\documentstyle[11pt,IAUS212,epsfig,twoside]{article}

\newcommand{\msun}{\mbox{$M_{\odot}$}}
\newcommand{\teff}{\mbox{$T_{\rm eff}$}}

\newcommand{\ratio}{\mbox{$v_{\infty}$/$v_{\rm esc}$}}
\newcommand{\mdot}{\mbox{$\dot{M}$}}
\newcommand{\reff}{\reference}

\def\edcomment#1{\iffalse\marginpar{\raggedright\sl#1\/}\else\relax\fi}
\reversemarginpar

\begin{document}
\vspace*{1cm}
\title{The winds of Luminous Blue Variables\\
 and the Mass of AG\,Car}
 \author{Jorick S. Vink$^1$, \& Alex de Koter$^2$}
\affil{$^1$Imperial College, London, UK\\
      $^2$University of Amsterdam, The Netherlands}
%\author{Co-Author}
%\affil{The Name of My Institution, The Full Address of My Institution}

\begin{abstract}
We present radiation-driven wind models for Luminous Blue 
Variables (LBVs) and predict their mass-loss rates. 
A comparison between our predictions and the observations of 
AG\,Car shows that the variable mass loss behaviour of LBVs 
is due the recombination/ionisation of Fe\,{\sc iv}/{\sc iii} and 
Fe\,{\sc iii}/{\sc ii}. We also derive a present-day mass of 35 \msun\
for AG\,Car.
\end{abstract}

\section{Introduction}

The strong winds of LBVs show a wide variety of mass-loss behaviour.
During their S\,Dor-type variations they expand in radius at 
approximately constant luminosity. In some cases the mass loss increases while the 
star expands (e.g. R\,71), whereas for 
others (e.g. R\,110) the behaviour is 
the exact opposite: as the star expands, its mass-loss rate drops. 
Recent radiation-driven wind models of OBA supergiants show that stars change 
their wind characteristics at spectral types B1 and A0, where \mdot\ jumps upwards
by factors of five, due to Fe recombinations. In this poster, we 
investigate whether these ``bi-stability jumps'' can also explain $\mdot(\teff)$ of LBVs.

\section{Mass Loss Predictions}

Our method is outlined in the poster by de Koter \& Vink (these proceedings). 
The full study is presented in Vink \& de Koter (2002).
Typical LBV results are shown in the first figure, for three values of \ratio.
It shows that there are ranges in $\teff$ where \mdot\ is predicted to 
increase, and ranges where \mdot\ decreases. 
The decreases are due to a growing mismatch between the 
positions of the driving lines (mostly in the UV) and the location of 
the flux maximum which shifts towards the optical for cooler stars.
The increases are due to recombinations of Fe\,{\sc iv} to Fe\,{\sc iii},
and Fe\,{\sc iii} to Fe\,{\sc ii}.
UV observations have shown that these ``bi-stability'' jumps 
occur at spectral types B1 and A0 (Lamers et al. 1995).

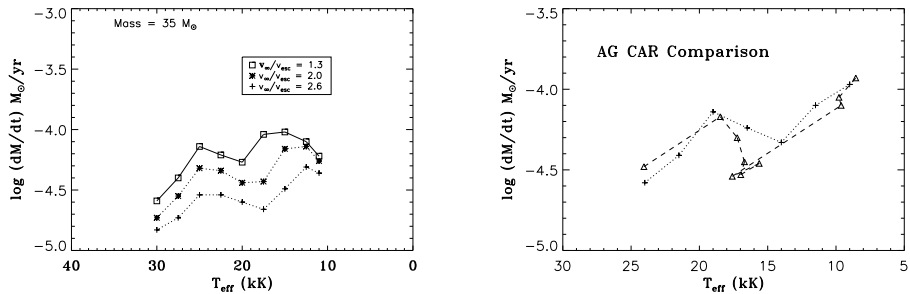
\begin{figure}
\plottwo{m35.ps.new}{agcar.ps.new}
\end{figure}

\section{Comparison with the observed Mass Loss behaviour of AG\,Car}

The second figure shows a comparison between our models (dotted line) and 
the H$\alpha$ mass loss rates for AG\,Car (Stahl et al. 2001).
It shows that the observed and predicted mass loss agree 
within 0.1 dex. Note that we have applied a corrective 
shift of $\Delta \teff = -6\,000$ K to our predictions to account for an 
inaccurate calculation of the ionisation balance of Fe.
The applied shift is consistent with constraints set by observations of supergiants 
which show that the jumps indeed occur at spectral types B1 and A0.
As \mdot(\teff) displays such a complex behaviour with fluctuations in 
\mdot\ of over more than 0.5 dex, this is a surprisingly good result. 
It shows that the mass-loss variability of AG\,Car is due to changes in
the ionisation balance of iron.

\section{The Mass of AG\,Car}

The predicted mass-loss rates are not only a function of \teff, but 
also of Mass. This sensitivity of \mdot\ to Mass offers 
the opportunity to constrain LBV Masses. 
Comparing the AG\, Car data with \mdot\ predictions 
for different masses results in a best fit to the present-day Mass of 
35\,\msun\ for AG\,Car. 

\section{Conclusions}

%The comparison of our predictions with the observed mass-loss 
%behaviour of LBVs has shown that:

\begin{itemize}

\item LBV winds are driven by radiation pressure.
\noindent
\item The mass loss behaviour of LBVs (of up to over 0.5 dex) 
      during their S\,Dor-type variation cycles can be explained 
      by the ionisation and recombination of 
      Fe\,{\sc iv}/{\sc iii} and Fe\,{\sc iii}/{\sc ii}.
noindent
\item The \mdot(\teff) behaviour of AG\,Car can be matched when 
      we adopt a mass of 35 \msun.

\end{itemize}

\end{document}